# Coarse–Super-Resolution–Fine Network (CoSF-Net): A Unified End-to-End Neural Network for 4D-MRI with Simultaneous Motion Estimation and Super-Resolution

Shaohua Zhi, Yinghui Wang, Haonan Xiao, Ti Bai, Hong Ge, Bing Li, Chenyang Liu, Wen Li, Tian Li and Jing Cai

**Abstract**: Four-dimensional magnetic resonance imaging (4D-MRI) is an emerging technique for tumor motion management in image-guided radiation therapy (IGRT). However, current 4D-MRI suffers from low spatial resolution and strong motion artifacts owing to the long acquisition time and patients' respiratory variations; these limitations, if not managed properly, can adversely affect treatment planning and delivery in IGRT. Herein, we developed a novel deep learning framework called the coarse–super-resolution–fine network (CoSF-Net) to achieve simultaneous motion estimation and super-resolution in a unified model. We designed CoSF-Net by fully excavating the inherent properties of 4D-MRI, with consideration of limited and imperfectly matched training datasets. We conducted extensive experiments on multiple real patient datasets to verify the feasibility and robustness of the developed network. Compared with existing networks and three state-of-the-art conventional algorithms, CoSF-Net not only accurately estimated the deformable vector fields between the respiratory phases of 4D-MRI but also simultaneously improved the spatial resolution of 4D-MRI with enhanced anatomic features, yielding 4D-MR images with high spatiotemporal resolution.

## 1 INTRODUCTION

Image-guided radiation therapy (IGRT) has been widely adopted in clinic for precision radiotherapy in patients with cancer [1]. In the past decade, magnetic resonance imaging (MRI) has gained much attention in IGRT because of its superior soft-tissue contrast and zero radiation hazard as compared to X-ray imaging techniques, such as computed tomography (CT) and cone-beam CT [2], [3]. In particular, MRI plays an important role in IGRT for abdominal cancers as it provides images with excellent anatomical details for accurate tumor volume delineation and possesses dynamic imaging capacity for tumor motion management [4], [5]. Respiratory motion can cause significant treatment errors if not managed properly. Thus, it is critical to manage respiratory motion when performing radiotherapy for abdominal cancers [6], [7]; this becomes particularly crucial when using stereotactic body radiation therapy (SBRT) [8], a modern radiotherapy technology that precisely delivers radiation treatment using a much higher radiation dose (10× higher) than conventional radiotherapy. Improper management of tumor motion when using SBRT can adversely affect patient treatment to a much greater degree than when it occurs in conventional radiotherapy.

Four-dimensional MRI (4D-MRI) is an emerging technique for motion management in the radiotherapy of mobile abdominal tumors. To date, various 4D-MRI techniques have been developed, and their promises have been well demonstrated [9]–[13]. One important imaging approach of 4D-MRI is fast volumetric MRI, in which the volume of interest is imaged at a sub-second speed, yielding real-time 4D-magnetic resonance (MR) images [14]. Furthermore, deformable image registration (DIR) can be performed on 4D-MR images to generate patient-specific motion models that depict voxel-wise motion patterns at different respiratory phases. 4D-MR images combined with the derived motion model are of great value in aiding precise radiotherapy, including 4D treatment planning, internal target volume determination, tumor tracking, 4D dose calculation, and organs at risk (OAR) sparing [15].

Currently, 4D-MRI is currently under investigation and development. There are a number of challenges to overcome before 4D-MRI can be fully adapted to the clinical setting. First, 4D-MR images suffer from limited spatial resolution; i.e., the temporal resolution of real-time 4D-MRI is approximately 1 s, whereas its voxel size is isotropically approximately 3 mm [9]. Owing to its insufficient image quality, as evidenced by a relatively low signal-to-noise ratio (SNR) and image artifacts caused by breathing variations, 4D-MRI may fail to display detailed anatomical structures. Second, it is challenging to calculate and model deformable vector fields (DVFs) from 4D-MR images for tumor tracking, primarily due to the extensive respiratory-related deformations and complicated soft anatomy variations in the abdominal region. These deficiencies of 4D-MRI can adversely affect its applications and diminish its values in IGRT.

Two potential solutions have been suggested to may help overcome the deficiencies of the current 4D-MRI techniques. First, super-resolution (SR) methods may help directly improve the MR image quality. Among the various SR methods, deep learning (DL)-based models are preferred for learning the mapping from low-resolution (LR) images to high-resolution (HR) images, thus restoring high-frequency structures as much as possible [16]–[19]. However, most MRI-related SR studies so far have focused on three-dimensional (3D)-MR images, while 4D-MR images remain under-explored. Second, the development of novel DIR methods is a more widely adopted strategy [20]–[22]. DL-based DIR models have been recently explored for 4D imaging [23]–[25]. The coarse-to-fine registration mechanism [26], [27] is particularly popular for DVF estimation; this mechanism predicts a rough DVF at a low image resolution and regards it as an initial guess for refinement in one or several steps at higher resolution levels.

Although DL-based models have resulted in breakthroughs, there is still much room for the improvement of 4D-MRI. For example, it is known that the accuracy of DVF estimation is highly sensitive to the quality of the images in DIR algorithms. For 4D-MRI, its voxel values at different respiratory phases may vary due to imaging factors and image artifacts. As a result, the compromised and inconsistent quality of 4D-MRI can lead to errors in DVF estimation. However, the relationship between motion modeling and image quality of 4D-MRI has not been thoroughly investigated. Although coarse-to-fine registration was proven successful for many medical imaging applications, it has limitations when applied to 4D-MRI, such as the loss of subtle structures and misalignments. Thus, in this study, we were motivated to develop an upgraded registration architecture to improve the performance of coarse-to-fine registration with respect to 4D-MR images.

The data preparation and preprocessing for 4D-MRI training also require careful consideration. First, obtaining the reference DVFs (training labels) for 4D-MRI registration is difficult, and the labeling process is time-consuming and laborious. Second, 4D-MRI is still under the investigational stage and is not being routinely used in clinical practice yet. 4D-MRI studies generally involve a small patient sample size (≤20 patients) [18], [28], [29], posing a great challenge for DL-based analysis because a small sample size can cause over-fitting during DL network training and subsequently affect model robustness. Second, it is common in radiotherapy that 3D T1-/T2-weighted MR images of the same patient are always available together with their 4D-MR images, which can be regarded as prior knowledge to promote image quality. It is clear that we need to develop DL networks tailored for 4D-MR images to overcome the mentioned limitations.

In this study, we aimed to achieve two goals by developing an end-to-end network capable of simultaneous accurate DVF estimation and image resolution enhancement. To this end, we developed a novel DL-based framework for 4D-MRI, also called the coarse–SR–fine network (CoSF-Net), by deeply excavating the inherent prior information on 4D-MRI. The main contributions of CoSF-Net are summarized as follows:

1) CoSF-Net encompasses three sub-models in an end-to-end fashion. To the best of our knowledge, it is the first DL framework capable of simultaneously enhancing the DIR and image quality for 4D-MRI.
2) An SR model was developed and embedded between the two registration submodels to construct a coarse–SR–fine architecture, which can boost the registration performance, especially when the input 4D-MRI pairs suffer from low spatial resolution [30]. In particular, a 2.5-dimensional conditional generative adversarial network (2.5D-cGAN) was designed to alleviate the issues of the limited number and imperfect matching of training pairs in 4D-MRI.
3) In both the coarse and fine DIR modules, the networks were trained in an unsupervised manner based on the VoxelMorph (VM) [24] model. Moreover, in the fine DIR model, we supplemented a feature extraction pathway for the prior MR image to strengthen detailed DVF estimation. Moreover, a residual DVF estimation mechanism was used for updating the refined DVF.
4) Extensive real-patient experiments were conducted with both visual comparison and quantitative evaluation to validate the effectiveness of CoSF-Net.

The rest of this article is organized as follows. In Section II, we describe the framework and implementation of the proposed network. Section III presents the experimental setup, data arrangement, evaluation metrics, and competitive algorithms adopted in this study. In Section IV, we report and analyze the results of our experiments conducted using real-patient data. In Section V, we discuss the relevant problems and conclude the study.

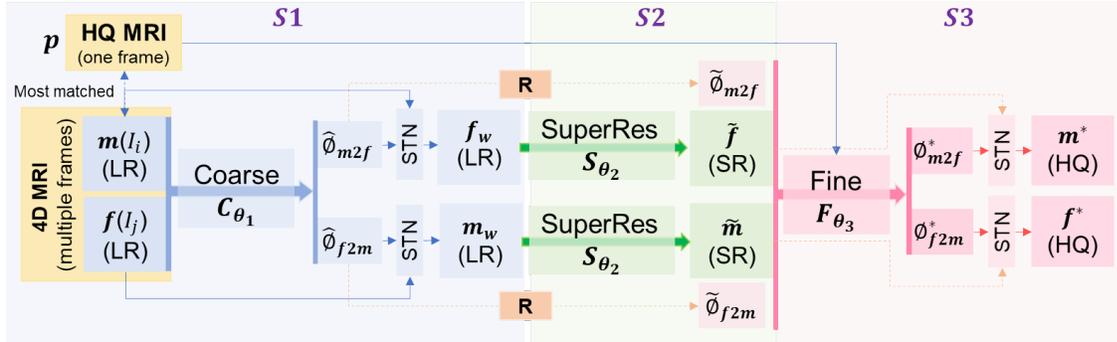

Fig. 1 A schematic illustration depicting the process of training CoSF-Net. The network involves the three cascaded submodels (S1, S2, and S3) in different colors: the coarse registration model ($C_{\theta_1}$, blue), the SR model ($S_{\theta_2}$, green), and the fine registration model ($F_{\theta_3}$, pink).

## 2 METHODS AND MATERIALS

### 2.1 Problem definition and notations

**Figure 1** presents a schematic illustration of the problem definition and notations of 4D-MRI. We denote $Q \supset \{I^1, I^2 ... I^K\}$, the original 4D-MRI sequence comprising **K** respiratory-correlated phases/frames. For simplicity, a pair of arbitrary phases $\{I^i, I^j\}$ ($I^i, I^j \in Q, i \neq j$) in 4D-MRI is denoted as the moving image $m$ and fixed image $f$, respectively. The purpose of our method is to develop a unified model $U_\theta$ parameterized by $\theta$ for enhancing the image quality of 4D-MRI while simultaneously estimating the DVF between the enhanced frames. The objective function can be described as follows:

$$\hat{\theta} = arg\min_\theta U_\theta(f, m, p), \quad (1)$$

where $p$ denotes a clinical T1-weighted MRI scan from the same patient, designated as a prior MR image. As shown in **Fig. 1**, the unified model is split into three cascaded submodels, including the coarse registration model ($C_{\theta_1}$, blue), the SR model ($S_{\theta_2}$, green), and the refined registration model ($F_{\theta_3}$, pink). The coarse DIR model $C_{\theta_1}$ is used to calculate the coarse DVF $\hat{\phi}$ using a DL-based model $\phi = C_{\theta_1}(f, m)$, considering the registration pair $\{f, m\}$ as the input. The optimization problem can be modeled as follows:

$$\hat{\theta_1} = arg\min_{\theta_1} \mathcal{L}_1\left(m, f, C_{\theta_1}(f, m)\right), \quad (2)$$

where $\theta_1$ denotes the learnable parameters of $C$ and $\mathcal{L}_1$ denotes the loss function.

We also denote the SR model $S_{\theta_2}$ parameterized by $\theta_2$ as a HR image ($I_{HR}$) generation procedure based on the observed LR counterpart ($I_{LR}$), the solution for which can be expressed by:

$$\hat{\theta_2} = arg\min_{\theta_2} \mathcal{L}_2(S_{\theta_2}(I_{LR}), I_{HR}), \quad (3)$$

where $S_{\theta_2}: I_{LR} \to I_{HR}$ can be replaced by the cGAN structure with a generator $G$ and discriminator $D$. Both $G$ and $D$ can be optimized in an alternative manner to solve the adversarial min-max problem as follows [1]:

$$S_{\theta_2} = \{G, D\}$$
$$G^* = arg\min_G \max_D \mathcal{L}_2(G, D). \quad (4)$$

Finally, the fine DIR model $F_{\theta_3}$ parameterized by $\theta_3$, feeds the enhanced HR image pair $\{\tilde{f}, \tilde{m}\}$ through $S_{\theta_2}$, the up-sampled DVF $\tilde{\phi} = R(\hat{\phi})$, together with the prior MR image $p$ to estimate a finer DVF $\phi^*$. The optimization function can be written as:

$$\hat{\theta_3} = arg\min_{\theta_3} \mathcal{L}_3\left(\tilde{f}, \tilde{m}, F_{\theta_3}(\tilde{f}, \tilde{m}, p, \tilde{\phi})\right) \quad (5)$$

To sum up, CoSF-Net is a cascade of three individual submodels; this can be denoted uniformly using the following equation: $U_\theta = \{C_{\theta_1}; S_{\theta_2}; F_{\theta_3}\}$.

### 2.2 Network architecture

The overall workflow of the proposed CoSF-Net is depicted in **Fig. 2 (a)** and can be outlined as follows. The first stage estimates a coarse DVF $\hat{\phi}$ of the input pair $\{f, m\}$ with a coarse DIR model $C_{\theta_1}$, deforming $\{f, m\}$ through a spatial transformation network (STN) to $\{f_w, m_w\}$. In the second stage, the HR images $\{\tilde{f}, \tilde{m}\}$ are recovered from $\{f_w, m_w\}$ through the SR model slice by slice. In the final stage, the recovered image pair $\{\tilde{f}, \tilde{m}\}$, the up-sampled DVF $\tilde{\phi}$, and the prior MR image $p$ are fed together into the fine DIR CNN $F_{\theta_3}$ to calculate an updated residual DVF $v$ and finally to obtain a finer DVF $\phi^*$ and the corresponding deformed HR images $\{f^*, m^*\}$. A combination of the three cascaded modules is considered the coarse–SR–fine structure.

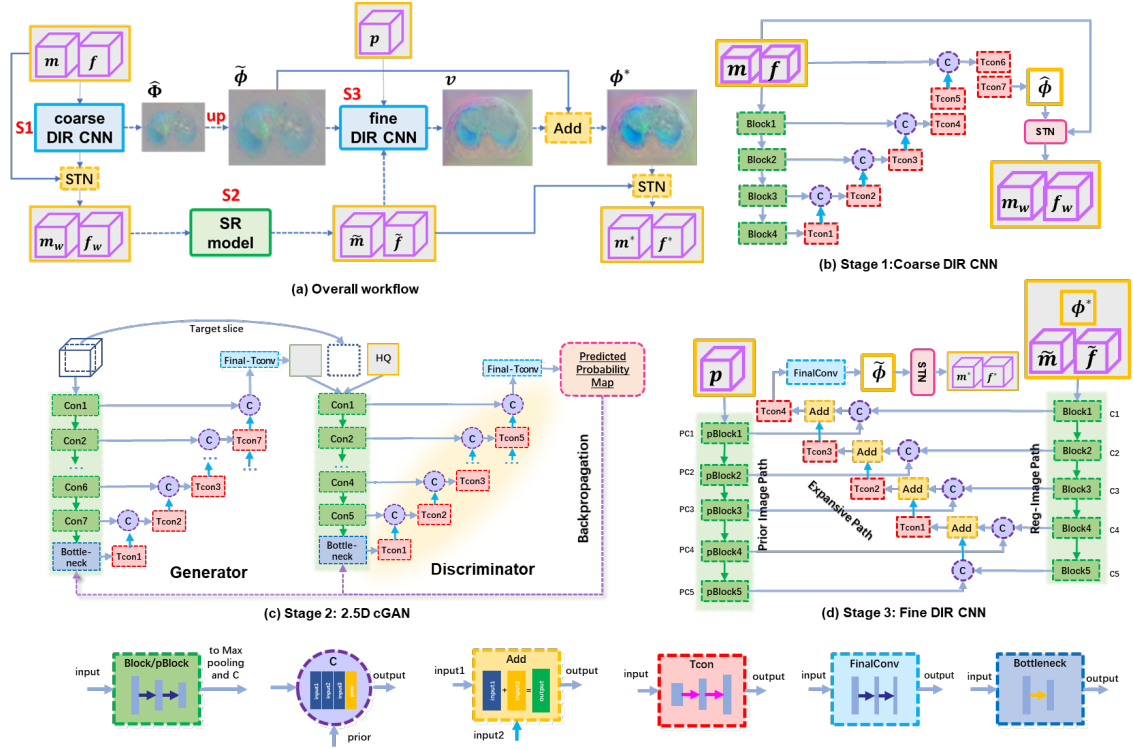

Fig. 2. The overall framework of the proposed network. (a) The workflow of CoSF-Net, which can be divided into (b), (c), and (d) three stages. (b) Stage 1: A coarse DIR CNN to predict a rough DVF between the phases in 4D-MRI at the low-resolution level; (c) Stage 2: A 2.5D-cGAN-based SR model to enhance the image quality and resolution of the deformed 4D-MR images; (d) Stage 3: A Fine DIR CNN to further compute the DVF residue for improved 4D-MR images from stage 2 at a high-resolution level, same as for the normal MRI.

### 2.2.1 Stage 1: Coarse DIR CNN

**Figure 2 (b)** shows the coarse DIR CNN, which inputs a concatenation of two arbitrary phases of MRI to predict the DVF between them. The network architecture was inspired by VM and trained in an unsupervised manner. To be specific, U-Net [2] comprises a contracting path of four 3D-convolutional blocks, a bottleneck connection, an expansive path of four 3D-convolutional blocks, and a final output layer. The STN [3] is used for calculating the deformed volume $M \circ \widehat{\phi}$, with the symbol $\circ$ denoting the deformable transformation operation based on the STN. In particular, we added an inverse-consistency penalty to render the DVF bidirectional $\widehat{\phi} = \{\widehat{\phi}_{m2f}, \widehat{\phi}_{f2m}\}$, which means that $f$ and $m$ can deform each other. Accordingly, the loss function in **Eq. (2)** contains two similarity terms: $\mathcal{L}_{sim}(.)$, measuring the image difference between the target and warped images, and a diffusion regularizer $\mathcal{L}_{smooth}(.)$, encouraging a smooth DVF.

$$\mathcal{L}_1\left(m, f, C_{\theta_1}(f, m)\right) = \mathcal{L}_{sim}\left(f, m \circ \widehat{\phi}_{m2f}\right) + \mathcal{L}_{sim}\left(m, f \circ \widehat{\phi}_{f2m}\right) + \lambda_1 \mathcal{L}_{smooth}(\widehat{\phi}), \qquad (6)$$

Negative normalized cross-correlation (NCC) was is employed in $\mathcal{L}_{sim}(.)$ instead of the L1 or L2 norm calculation owing to the varying intensities among phases in 4D-MRI. The regularization parameter $\lambda_1$ is used to control the trade-off between the fidelity and regularization terms.

### 2.2.2 Stage 2: SR Network

In stage 2, the SR model uses cGANs to synthesize improved MR images from the 4D-MR image counterparts. **Figure 2(c)** illustrates the detailed architecture of the SR model, also known as the 2.5D-cGAN. The proposed model inputs the 2.5D images by considering the presence of limited volumetric MRI training data and the high inter-slice correlation within the volume. Hence, five consecutive 2D transversal slices are intergraded and fed into the 2.5D-cGAN for predicting the central slice. Apart from the pixel-wise L1-based intensity loss ($\mathcal{L}_{L1}(G)$) and an adversarial loss ($\mathcal{L}_{GAN}(G,D)$), a multi-scale structural similarity index (MS-SSIM) ($\mathcal{L}_{SSIM}(G)$) [4] is incorporated into the total loss function of the 2.5D-cGAN, restoring structural information effectively as follows:

$$\mathcal{L}_2(G,D) = \mathcal{L}_{GAN}(G,D) + \lambda_2 \mathcal{L}_{L1}(G) + \lambda_3 \mathcal{L}_{SSIM}(G)$$
$$= \mathbb{E}_{(I_{LR},I_{SR})}[\log D(I_{LR},I_{SR})] + \mathbb{E}_{I_{LR}}\left[\log\left(1 - D(I_{LR},G(I_{LR}))\right)\right]$$
$$+ \lambda_2 \mathbb{E}_{(I_{LR},I_{SR})}[\|I_{SR} - G(I_{LR})\|_1] + \lambda_3(1 - SSIM(I_{SR},D(I_{SR}))) \quad (7)$$

where $\mathbb{E}(.)$ indicates the expected value, and the MS-SSIM index is computed using the same parameters proposed in a previous study [5]. For technical details, the generator $G$ comprises seven blocks of convolution-BN-ReLU operations in the encoder and decoder, while the discriminator $D$ contains five identical blocks.

### 2.2.3 Stage 3: Fine DIR CNN

As mentioned in **Section II A**, a combination of the three submodels can be regarded as a coarse–SR–fine structure. The architecture of the fine DIR CNN is displayed in **Fig. 2(d)**. It contains two independent paths in the encoder module for extracting multi-level features; one is called Reg-Image-Path for the registration pair $\{\tilde{f}, \tilde{m}\}$, whereas the other is called Prior-Image-Path for $p$. A concatenation of both features at the same scale is then delivered into the decoder module. It is worth noting that the prior MR image $p$ is not perfectly matched with the moving image. Before integrating $p$ into the network, it was pre-aligned to be in the same phase as the enhanced moving image $\tilde{m}$. Instead of explicitly calculating $\phi^*$, we adopted a residual DVF calculation strategy. This strategy utilizes $\tilde{\phi}$ as an initial guess to obtain a warped volume using the function $\tilde{m} \circ \tilde{\phi}_{m2f}$. For back-propagation, the loss function of the fine DIR CNN is accordingly modified as follows:

$$\mathcal{L}_3\left(\tilde{f},\tilde{m},F_{\theta_3}(\tilde{f},\tilde{m},p,\tilde{\phi})\right) = \alpha \mathcal{L}_{sim}(\tilde{f},\tilde{m} \circ \tilde{\phi}_{m2f}) + (1-\alpha)\mathcal{L}_{sim}(\tilde{f},p \circ \tilde{\phi}_{m2f}) + \lambda_4 \mathcal{L}_{smooth}(\tilde{\phi}) \quad (8)$$

where the predicted $\tilde{\phi}_{m2f}$ is constrained by two similarity forms: one is controlled by $\tilde{f}$ with improved resolution, whereas the other one is attributed to the prior MRI. The parameter $\alpha$ is adopted to adjust the contribution factors of both similarity metrics. The $\mathcal{L}_{sim}(.)$ and $\mathcal{L}_{smooth}(.)$ in **Eq. (8)** for fine DIR CNN are the same as those in **Eq. (6)** for coarse DIR CNN. Using the proposed fine DIR CNN, we could then estimate a residual DVF $v$ between $\tilde{m} \circ \tilde{\phi}_{m2f}$ and $\tilde{f}$. $v$ not only reflects a more accurate DVF but also contains detailed anatomic changes. In doing so, the updated DVF can be obtained with $\phi^* = v + \tilde{\phi}$.

## 3  EXPERIMENTS

### 3.1  Dataset Preparation

The MRI data used in this study were acquired from patients with liver tumors undergoing radiotherapy using a 3T scanner (Skyra, Siemens, Erlangen, Germany). The study protocol was approved by the institutional review board. The patients underwent regular 3D MRI scans (T1- and T2-weighted), which were designated as "prior MRIs." In addition to regular 3D MRI, each patient also underwent 4D-MRI using the TWIST volumetric interpolated breath-hold examination (TWIST-VIBE) MRI sequence, continuously generating 72 frames covering several respiratory cycles. Ten frames covering a breathing cycle were then selected from the original 4D-MRI using the body area method [6]; these frames represent the respiratory-correlated phases used in this study. The dimension of each volume of the 4D-MRI is $160 \times 128 \times 64$, with a voxel size of $2.7 \times 2.7 \times 3.0$; the dimensions of the prior MRI is $320 \times 320 \times 72$, with a smaller voxel size of $1.2 \times 1.2 \times 3.0$.

For network training and evaluation, we retrospectively included a total of twenty-seven MRI patients, all of whom had a 4D-MRI scan and prior 3D MRI scans. Twenty patients were included for network training and validation, and the remaining seven patients were included for testing. To address the issue of limited training data, we performed data augmentation by expanding the dataset by 90°–180°–270° of rotation. This yielded 2,480 2D pairs of transverse MR images for the 2.5D-cGAN training. Moreover, we set the respiratory phase number to 10 and quantified the degree of deformation change into 4 grades (named as phase range) according to the breathing amplitude for further data augmentation. Finally, a total of 3,200 volumetric pairs were obtained for DIR CNN training. Note that the intensity of all the images was normalized to 0–1; 90% of the 20 patients were randomly chosen for training, whereas the remaining 10% were assigned for validation. Regarding the arrangement of the seven testing datasets, four representative cases were displayed for visualization, and all seven patients were include d in the quantitative analysis.

### 3.2  Implementation details

**Training procedure/strategy:** To ensure that the three submodels in CoSF-Net play their expected roles, we trained the framework in two stages. First, the coarse DIR CNN and 2.5D-cGAN were pre-trained individually to initialize the filters in CNNs. Second, the fine DIR CNN was cascaded with the two pre-trained submodels, and CoSF-Net was jointly tuned in an end-to-end manner, ensuring a minimum and stable output.

**Hyper-parameter setting:** A total of 500 epochs were obtained for the coarse DIR CNN. The parameter $\lambda_1$ in **Eq. (6)** was set as 4 to ensure a trade-off performance. In the 2.5D-cGAN training process, the network converged after 500 epochs with a batch size of 15. We set $\lambda_2 = 10$ and $\lambda_3 = 10$ in **Eq. (7)**. The learning rates of these two submodels were set to 4e-5 initially and decreased to 90% after every 30 epochs. Finally, in the CoSF-Net tuning procedure, the initial learning rate was set to 5e-5 and adjusted to 90% after every 10 epochs. Meanwhile, the hyper-parameter was set as $\alpha = 0.35$ and the regularization parameter was set as $\lambda_4 = 5$ in **Eq. (8)** empirically.

The proposed CoSF-Net was implemented on an NVIDIA GTX3090 GPU in the PyTorch framework. The corresponding training and testing codes will be available on the authors' website once the paper is published.

*3.3 Model Evaluation*

We evaluated the proposed CoSF-Net both qualitatively and quantitatively in the following aspects: 1) We analyzed the intermediate results at individual stages to evaluate the effectiveness of the proposed coarse-SR-fine framework; 2) We compared CoSF-Net with existing DL-based neural networks and conventional optimization-based algorithms; 3) We conducted an ablation study to investigate the impact of the designed components on the network performance; 4) We analyzed tumor localization and feature recovery in 4D-MR images using the proposed CoSF-Net.

*3.3.1 Comparison with existing algorithms*

To evaluate the performance of the proposed method, we compared it with several state-of-the-art methods. In terms of the registration ability, we compared the proposed network with the following three classical registration algorithms: pTV algorithm [7], Elastix [8], and Demons [9]. Furthermore, two DL-based methods were employed to assess the effectiveness of CoSF-Net, including the supervised VM (sVM) and single-scale unsupervised VM (uVM) models. In implementing the supervised VM, we adopted the pTV algorithm for generating reference DVFs for network training. For the SR recovery ability, we used the enhanced deep SR network (EDSR) model [10] for comparison. All the parameters were carefully selected to ensure a fair comparison.

*3.3.2 Evaluation metrics*

For quantitative evaluation, we measured the error distance between the warped and fixed images using the rooted mean square error (RMSE), the structural similarity index metric (SSIM) [11], the peak signal-to-noise ratio (PSNR), and normalized mutual information (NMI) [12], respectively. These evaluation metrics reflect the model performance in different aspects, such as RMSE calculates the absolute difference between the restored image and the ground truth; the structural similarity index metric evaluates the preserving ability of structural information; the PSNR represents the noise suppression ability; and the NMI evaluates the correlation between the ground truth and the reconstructed images. A larger NMI value indicates a higher similarity with the ground truth, whereas a smaller value indicates a lower similarity.

## 4 RESULTS

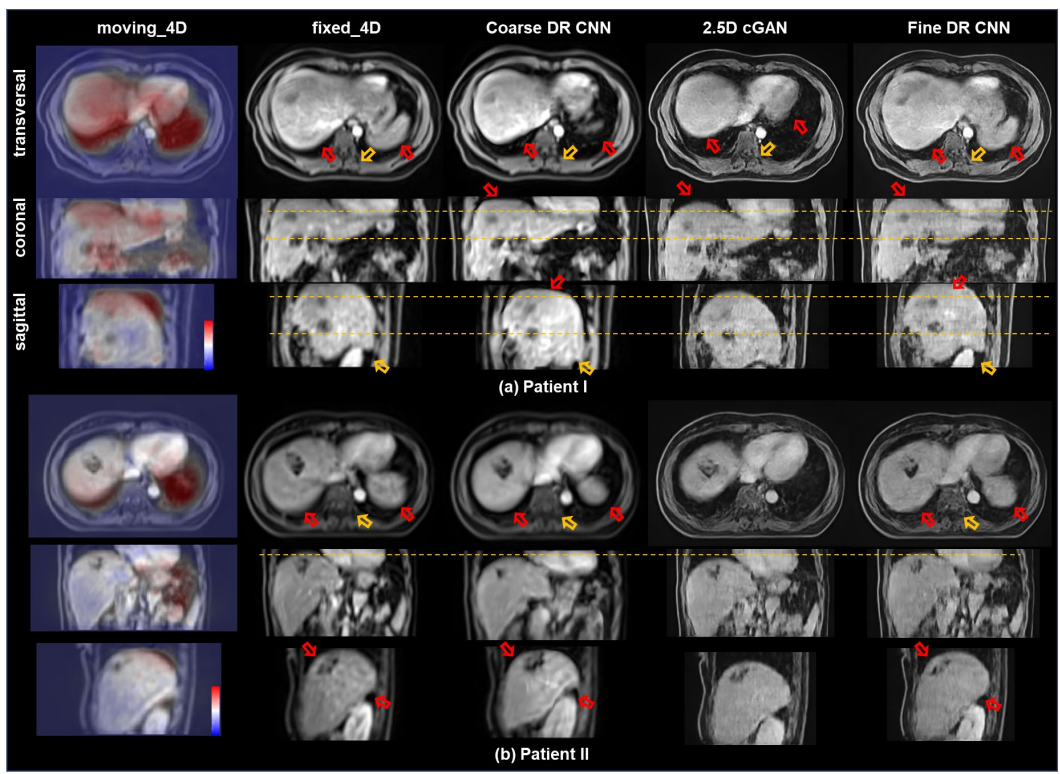

Fig. 3 The results of 4D-MRI for (a) Patient I and (b) Patient II. From left to right: moving image in original 4D-MRI (moving_4D), fixed image in original 4D-MRI (fixed_4D), intermediate warped image from the moving image by coarse DIR CNN, enhanced moving image via 2.5D-cGAN, and final warped image by fine DIR CNN. Especially, the color-coded DVF predicted by CoSF-Net overlaps the moving_4D, showing the variation degree of DVF. All the volumes are displayed in the transversal, coronal, and sagittal views. The display window is provided at a grayscale window of C = 0.4, W = 0.8.

### 4.1 Intermediate Results Analysis

**Figure 3** presents the intermediate results for Patient I (a) and II (b). Particularly, we selected the registration pair with the maximum phase range to test the robustness of the network; that is, the moving image is at the end-of-inhale (EOI) phase (first column), whereas the fixed image is at the end-of-exhale (EOE) phase (second column). As observed in **Fig. 3**, considering the LR image pair as the input, the warped volume by coarse DIR CNN (third column) is similar to that of the fixed image in the second column. However, misalignments still exist, such as that in the shape of the diaphragm (right arrows in the three visual views), which indicates that the DVF by coarse DIR CNN roughly depicts the contour changes of the registration pair but fails to estimate the detailed deformations. Using the SR model, the image quality of the generated image has improved; the artifacts are suppressed and small structures and sharp contour shapes are recovered, as compared with the original 4D-MR images obtained at the EOI phase. Finally, by feeding the initial DVF using a coarse DIR CNN and an image pair with better resolution, the fine DIR CNN could predict a more accurate DVF (red arrows) and retain more detailed anatomic structures (yellow arrows) in the resulting images of both patients. In addition to the intermediate results, a moving image with a color-coded updated DVF is displayed in the first column to show the deformation changes, which are discussed in Section IV.

### 4.2 Ablation study

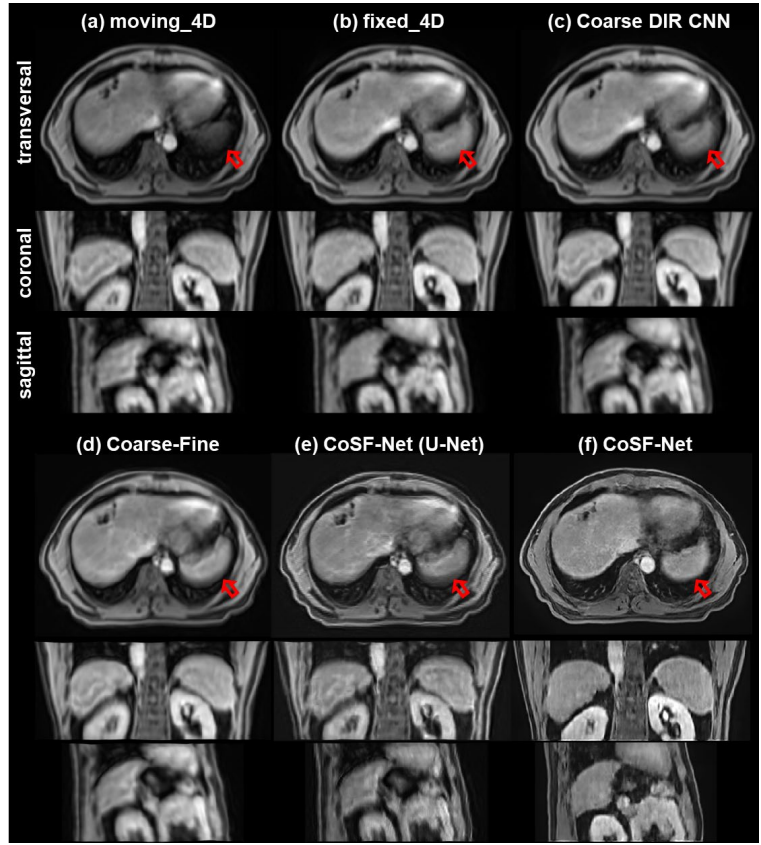

Fig. 4 Visual comparison of Patient III in the three views for investigating the contribution of the SR model and a prior MRI component in the architecture of CoSF-Net. From left to right: (a) and (b) represent the original registration image pair in 4D-MRI; (c) the warped image by sole coarse DIR CNN; (d) the results by classical coarse-to-fine model; (e) the results by CoSF-Net without a prior MRI; and (f) the results by the complete CoSF-Net.

In the ablation study, we investigated the effect of two components in CoSF-Net on the final performance, which are the participation of the SR model in coarse-SR-fine workflow and the incorporation of the prior MRI component in the fine DIR CNN. **Fig. 4** presents the results for the case of patient III. After removing both the SR model and the prior MRI, the remaining structure was found to be a classical coarse-to-fine architecture, referred to as coarse–fine in **Fig. 4(d)**. As a result, the image quality in **Fig. 4(d)** has not yet been improved, although it obtains a better DVF estimation than the pure coarse DIR CNN in **Fig. 4(c)**. Furthermore, as shown in **Fig. 4(e)**, the fine DIR CNN is simplified to a U-Net architecture without the prior MRI component, referred to as CoSF-Net (U-Net). It can be seen that the image quality improves to some extent, but some detailed structures in the warped image appear unrealistic. Compared with **Fig. 4(d) and (e)**, the image generated by the complete CoSF-Net showed superior performance, indicating that the SR model and prior MRI help preserve the anatomic features and topology of real patients.

### 4.3 Comparison with Existing Methods

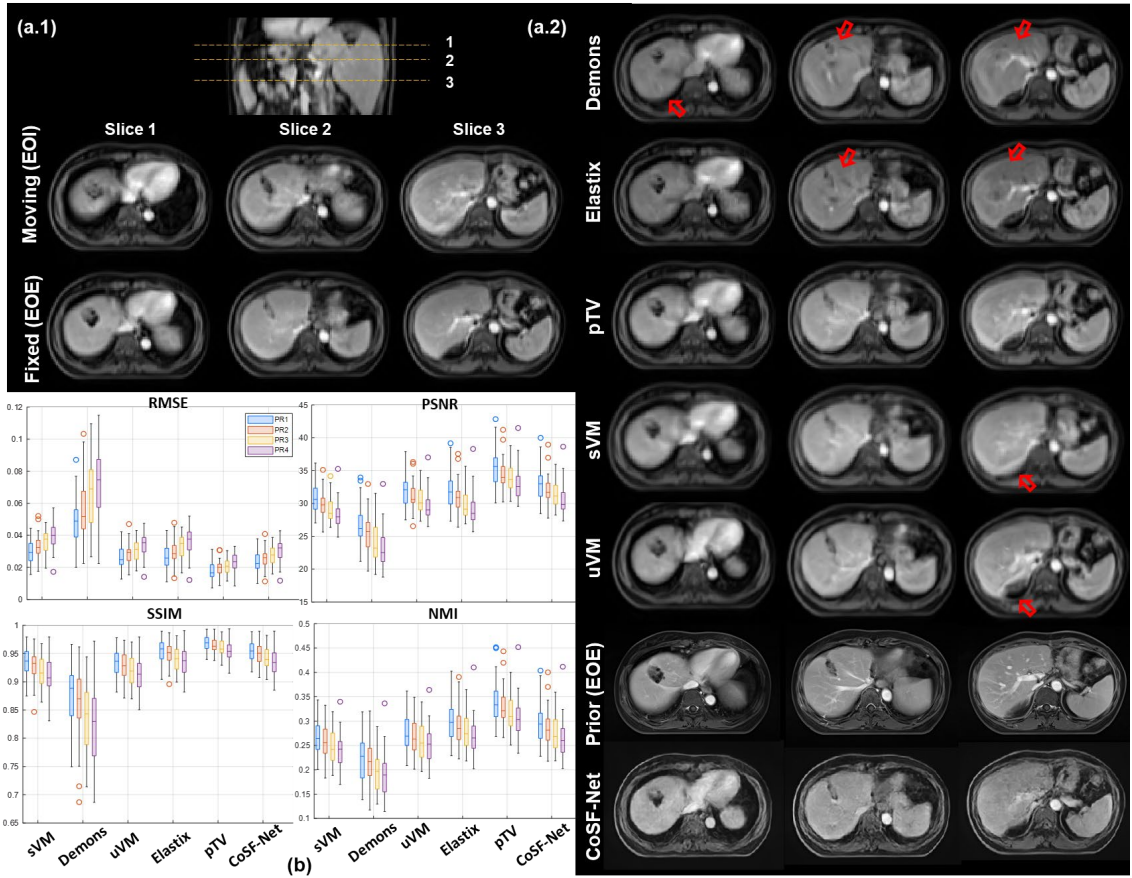

Fig. 5 Visual results and statistical analysis results for Patient I. Sub-figure (a.1): The registration pair of original 4D-MRI. Sub-figure (a.2): Deformation results obtained by warping the moving image by the predicted DVF using the following models: Demons, Elastix, pTV, supervised VM, coarse DIR CNN, prior MRI, and CoSF-Net. Three individual slices were selected in the transversal view for comparison. Sub-figure (b): Four evaluation metrics (RMSE, PSNR, SSIM, and NMI) were calculated in different phase ranges as represented by the blue, red, yellow, and purple box plots, respectively. Note that the moving image used in this experiment is up-sampled directly without using the SR model. All the images are displayed in the window of level C = 0.4 and window W= 0.8.

Figure 5 (a) shows the qualitative results for Patient II using CoSF-Net and other methods for comparison. Three slices profiled along the superior-inferior dimension in the same respiratory phase were chosen for visualization. In terms of the quantitative analysis, **Fig. 5(b)** displays the results of the different phase ranges (PR = 1–4) were also involved in testing the robustness of the models.

As shown in **Fig. 5(a.2)**, the results obtained by Demons and Elastix showed unsatisfactory deformation. Both methods introduce inaccurate features or fail to maintain consistent intensities, as indicated by the red arrows, probably because both methods are sensitive to distorted image quality. The results using pTV maintain the correct physical topology and outperform those obtained using Demons and Elastix in terms of four metrics, as illustrated in **Fig. 5(b)**. As for the DL-based DIR methods, compared with sVM, the sole coarse DIR CNN is limited in estimating enough accurate DVF due to its self-learning property. The bottom two rows display the deformed image from EOI to EOE using DVF predicted by CoSF-Net and the resultant image by CoSF-Net, respectively. Overall, CoSF-Net could predict an excellent 4D-MR image with accurate DVF estimation and high image quality. In the quantitative evaluation, there are no ground truth images for the real patient dataset. To obtain a fair comparison, we used the fixed image (EOE) as the baseline for comparative methods. Moreover, we used prior MRIs (EOE) as the baseline for calculating the metrics for CoSF-Net. We demonstrate that the registration performance of CoSF-Net is better than that of the other DL-based models and ranks second in all four quantitative metrics [**Fig. 5(b)**]; the pTV algorithm displayed the best performance, and its results were just slightly better than those of CoSF-Net. Besides, benefiting from the DIR plus SR mechanism, CoSF-Net achieves a considerable quality improvement in detailed feature recovery, whereas the other models fail.

### 4.4 Analysis of Tumor Localization and Detailed Structure Recovery Ability

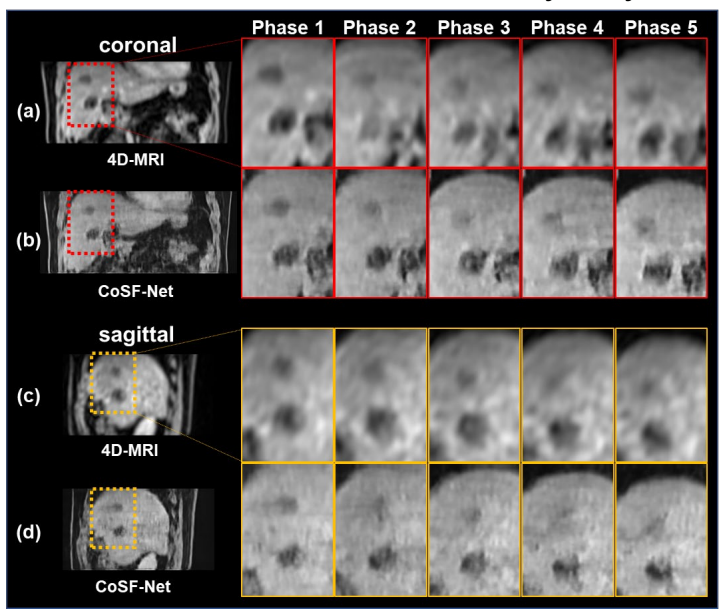

Fig. 6 Analysis of tumor localization and motion trajectory for patient I. The transversal and coronal views of the original 4D-MR images are shown in the first (a) and third (c) rows, respectively, whereas the CoSF-Net equivalents are shown in the second (b) and fourth (d) rows. To identify the tumor location and shape, two

ROIs were chosen in the coronal (red rectangle) and sagittal (yellow rectangle) views. On the right side of the figure, successive zoom-in images from the EOE to EOI phases are shown. The display window is C = 0.35 and W = 0.70.

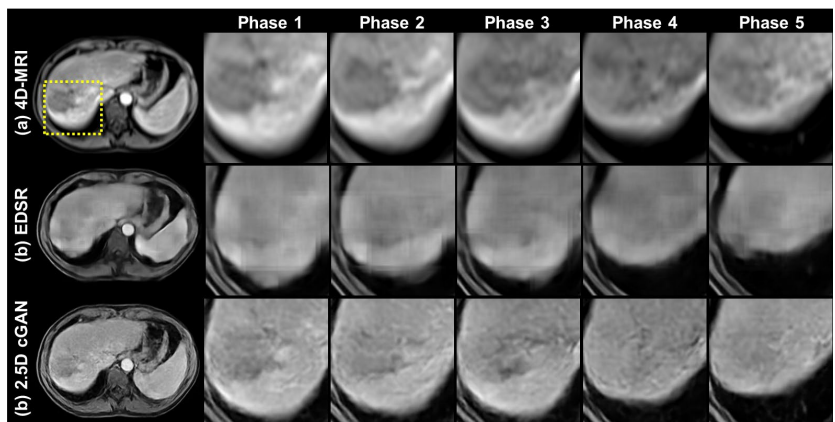

Fig. 7 Anatomical feature recovery results for Patient IV: (a) Original 4D-MR images, (b) images obtained using the EDSR, (c) images obtained using 2.5D-cGAN. An ROI is selected in the transversal view (yellow rectangle) to represent the detailed features affected by respiratory motion. The display window is provided at a grayscale window of C = 0.4 and W = 0.7.

To demonstrate the effectiveness of tumor localization and feature recovery via CoSF-Net, we chose two patient cases (Patient I and IV) with apparent tumors close to the diaphragm; these tumors move with respiratory motion. In the two different visual views presented in **Fig. 6**, the tumors and diaphragm margin are indicated with red and yellow rectangles, respectively. Since real-patient datasets lack a HR ground truth, we used the original LR 4D-MR images as the baseline [**Fig. 6(a)(c)**]. Compared with the original 4D-MR images, the resulting ROIs obtained using CoSF-Net at various phases yield an improved contour depiction of the tumor and diaphragm while maintaining an accurate motion trajectory.

Five consecutive phases of zoom-in patches (yellow rectangles) are shown in **Fig. 7** for illustrating the detailed feature recovery performance of the proposed 2.5D-cGAN and its competitors. Although the original 4D-MR images display the tumor's motion trajectory near the diaphragm, it suffered from LR and poor PSNR. Applying EDSR to the 4D-MR images effectively reduced the noise; however, the anatomical features could not be recovered; over-smoothing is also observed because EDSR does not account for the non-uniqueness property of 4D-MRI. Unlike EDSR, the proposed 2.5D-cGAN restores tiny structures to the maximum possible extent, providing a satisfactory spatiotemporal resolution.

## 5  DISCUSSIONS AND CONCLUSION

In this study, we developed a unified DL framework to reconstruct a sequence of 4D-MR images with enhanced spatiotemporal resolution and image quality for application in radiotherapy. The proposed model enables simultaneous motion estimation and image resolution enhancement in 4D-MR images using a cascade of three submodels, including a coarse DIR CNN, an SR model, and a fine DIR CNN. Extensive experiments have been conducted to demonstrate that the proposed CoSF-Net can predict accurate DVFs between the respiratory phases of 4D-MR images and effectively enhance the image resolution. Ultimately, the 4D-MR images generated using CoSF-Net have a much higher spatial resolution and depict more detailed anatomical features than the original 4D-MR images.

As demonstrated in **Fig. 3** and **Fig. 4(c) and (d),** a single DIR model or coarse-to-fine workflow alone cannot simultaneously predict the DVFs and yield high-quality images. Based on the abovementioned observations, we developed the CoSF-Net to achieve two tasks using one unified model. Ablation experiments [**Fig. 4(c)-(e)**] were conducted to verify the reasonability and necessity of the coarse–SR–fine mechanism.

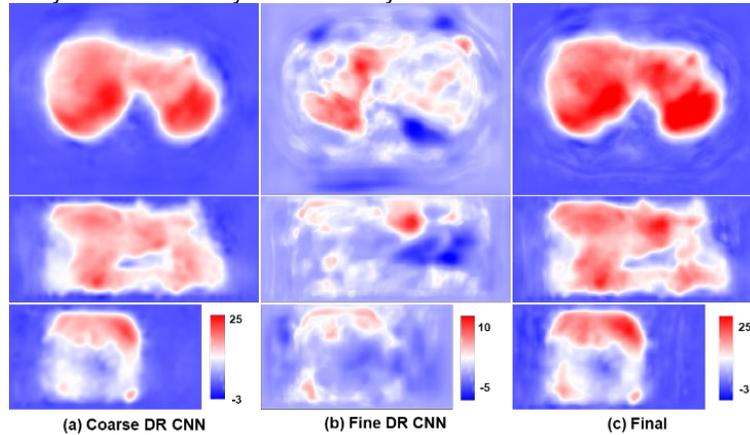

Fig. 8 A heatmap of estimated DVF for Patient I at different stages. (a) Coarse DR CNN: the estimated DVF by coarse DR CNN; (b) fine DR CNN: the estimated residue DVF by fine DR CNN; (c) final: final DVF estimation by adding (a) and (b). The magnitude of DVF is shown in color; red indicates a larger transformation, whereas blue color indicates that the volume is static or has smaller movements.

We note that improvement of the 4D-MR image quality increased the accuracy of DVF estimation, as shown in the intermediate results depicted in **Fig. 3**. To the best of our knowledge, CoSF-Net is the first DL model that considers both motion estimation and image enhancement in 4D-MRI, and no such model for 4D-MRI exists thus far.

In CoSF-Net, a cascade of three submodels, each of which is essential and indispensable, is well-conceived to comprehensively exploit the inherent image dynamics of 4D-MRI. We derive the following considerations concerning the design of CoSF-Net. First, we built a GAN-based SR model. Unlike traditional medical imaging SR tasks, the training MRI pairs exhibit varying intensity distributions and breathing-related motion patterns [13]. Thus, conventional CNNs with L-1 or L-2 norm-based loss functions may be inappropriate for 4D-MRI reconstruction. Benefiting from the mechanism of the discriminator module, GANs help the network to learn the dependence (perceptual loss) between image pairs, rendering the GAN results more consistent with the actual perception of human vision. The comparison results presented in **Fig. 7** indicate that GAN is a suitable technique for restoring the detailed features of 4D-MRI. Second, the results of our previous studies show that including prior images in both classical algorithms and DL networks can improve the 4D-cone bean CT image quality [12], [14], [15]. However, in this study, the prior MR image is used to facilitate the estimation of DVF. Moreover, the prior MR image is used in both architecture design and back-propagation calculation, as described in **Section II.C-(3)**. Hence, CoSF-Net outperformed the other controlled settings when guided by prior MRI, as illustrated in **Fig. 4 and Table I**. Furthermore, as described in **Section II.B-(3)**, we calculated the residual DVF to refine the updated DVF instead of making a straightforward DVF prediction based on residual flow networks [16]. **Figure 8** presents the heatmaps of DVF generated at different stages in CoSF-Net. By estimating the residual DVF, we found that the final DVF, as shown in **Fig. 8(c),** can effectively excavate more detailed information than the predicted DVF through coarse DR CNN [**Fig. 8(a)**].

We also draw up the future prospects for our study as follows. First, the through-plan resolution of enhanced 4D-MR images can be further improved. As we know that clinical MRI usually has a lower through-plan resolution than in-plane resolution. We will upgrade our SR model to achieve isotropic HR in our future study. Second, besides anatomical imaging, MRI can perform functional imaging, making it a powerful tool for treatment response assessment and outcome prediction. This promotes us to develop a contrast-variant 4D-MRI (5D) model based on the currently proposed 4D-MRI network. One possible approach is to extend the current pair-wise registration to a group-wise registration by introducing the temporal correlation property in network design, perhaps by using long short-term memory (LSTM)-based architecture [17].

In conclusion, we proposed an innovative DL model capable of simultaneous motion modeling and resolution enhancement for 4D-MRI. The proposed CoSF-Net integrates a GAN-based SR model into the coarse-to-fine registration model and presents a coarse–SR–fine framework. We also upgraded the model by considering prior knowledge and limited 4D-MR image datasets. Our results obtained using a wide range of real patient datasets showed that CoSF-Net can handle motion estimation and image resolution enhancement in a unified model. Moreover, CoSF-Net was shown to successfully recover 4D-MR images with a better spatiotemporal resolution than other state-of-the-art networks and algorithms.

**REFERENCES**

[1] L. Xing, B. Thorndyke, E. Schreibmann, Y. Yang, T.-F. Li, G.-Y. Kim, G. Luxton, and A. Koong, "Overview of image-guided radiation therapy," Medical Dosimetry, vol. 31, no. 2, pp. 91–112, Jun. 2006.

[2] M. J. Menten, A. Wetscherek, and M. F. Fast, "MRI-guided lung SBRT: Present and future developments," Physica Medica, vol. 44, pp. 139–149, Dec. 2017.

[3] R. Otazo, P. Lambin, J.-P. Pignol, M. E. Ladd, H.-P. Schlemmer, M. Baumann, and H. Hricak, "MRI-guided radiation therapy: An emerging paradigm in adaptive radiation oncology," Radiology, vol. 298, no. 2, pp. 248–260, Feb. 2021.

[4] J. Cai, G. W. Miller, T. A. Altes, P. W. Read, S. H. Benedict, E. E. de Lange, G. D. Cates, J. R. Brookeman, J. P. Mugler, and K. Sheng, "Direct measurement of lung motion using hyperpolarized helium-3 MR tagging," Int J Radiat Oncol Biol Phys, vol. 68, no. 3, pp. 650–653, Jul. 2007.

[5] J. Cai, P. W. Read, J. M. Larner, D. R. Jones, S. H. Benedict, and K. Sheng, "Reproducibility of interfraction lung motion probability distribution function using dynamic MRI: statistical analysis," Int J Radiat Oncol Biol Phys, vol. 72, no. 4, pp. 1228–1235, Nov. 2008.

[6] H. Ge, J. Cai, C. R. Kelsey, and F.-F. Yin, "Quantification and minimization of uncertainties of internal target volume for stereotactic body radiation therapy of lung cancer," Int J Radiat Oncol Biol Phys, vol. 85, no. 2, pp. 438–443, Feb. 2013.

[7] I. Vergalasova and J. Cai, "A modern review of the uncertainties in volumetric imaging of respiratory-induced target motion in lung radiotherapy," Medical Physics, vol. 47, no. 10, pp. e988–e1008, 2020.

[8] M. Fast, A. van de Schoot, T. van de Lindt, C. Carbaat, U. van der Heide, and J.-J. Sonke, "Tumor trailing for liver SBRT on the MR-Linac," Int J Radiat Oncol Biol Phys, vol. 103, no. 2, pp. 468–478, Feb. 2019.

[9] J. Cai, Z. Chang, Z. Wang, W. Paul Segars, and F.-F. Yin, "Four-dimensional magnetic resonance imaging (4D-MRI) using image-based respiratory surrogate: A feasibility study," Med Phys, vol. 38, no. 12, pp. 6384–6394, Dec. 2011.

[10] Y. Liu, F.-F. Yin, B. G. Czito, M. R. Bashir, and J. Cai, "T2-weighted four dimensional magnetic resonance imaging with result-driven phase sorting," Med Phys, vol. 42, no. 8, pp. 4460–4471, Aug. 2015.

[11] Y. Liu, X. Zhong, B. G. Czito, M. Palta, M. R. Bashir, B. M. Dale, F.-F. Yin, and J. Cai, "Four-dimensional diffusion-weighted MR imaging (4D-DWI): a feasibility study," Med Phys, vol. 44, no. 2, pp. 397–406, Feb. 2017.

[12] W. Harris, F.-F. Yin, J. Cai, and L. Ren, "Volumetric cine magnetic resonance imaging (VC-MRI) using motion modeling, free-form deformation and multi-slice undersampled 2D cine MRI reconstructed with spatio-temporal low-rank decomposition," Quant Imaging Med Surg, vol. 10, no. 2, pp. 432–450, Feb. 2020.

[13] T. Li, D. Cui, E. S. Hui, and J. Cai, "Time-resolved magnetic resonance fingerprinting for radiotherapy motion management," Med Phys, vol. 47, no. 12, pp. 6286–6293, Dec. 2020.

[14] J. Yuan, O. L. Wong, Y. Zhou, K. Y. Chueng, and S. K. Yu, "A fast volumetric 4D-MRI with sub-second frame rate for abdominal motion monitoring and characterization in MRI-guided radiotherapy," Quant Imaging Med Surg, vol. 9, no. 7, pp. 1303–1314, Jul. 2019.

[15] C. Wang and F.-F. Yin, "4D-MRI in radiotherapy," in Magnetic Resonance Imaging, L. Manchev, Ed. IntechOpen, 2019.

[16] B. Lim, S. Son, H. Kim, S. Nah, and K. M. Lee, "Enhanced deep residual networks for single image super-resolution," arXiv:1707.02921 [cs], Jul. 2017.

[17] X. Du and Y. He, "Gradient-guided convolutional neural network for MRI image super-resolution," Applied Sciences, vol. 9, no. 22, p. 4874, Jan. 2019.

[18] S. Park, H. M. Gach, S. Kim, S. J. Lee, and Y. Motai, "Autoencoder-inspired convolutional network-based super-resolution method in MRI," IEEE Journal of Translational Engineering in Health and Medicine, vol. 9, pp. 1–13, Apr. 2021.

[19] C. Zhao, B. E. Dewey, D. L. Pham, P. A. Calabresi, D. S. Reich, and J. L. Prince, "SMORE: A Self-Supervised Anti-Aliasing and Super-Resolution Algorithm for MRI Using Deep Learning," IEEE Transactions on Medical Imaging, vol. 40, no. 3, pp. 805–817, Mar. 2021.

[20] V. Vishnevskiy, T. Gass, G. Szekely, C. Tanner, and O. Goksel, "Isotropic total variation regularization of displacements in parametric image registration," IEEE Transactions on Medical Imaging, vol. 36, no. 2, pp. 385–395, Feb. 2017.

[21] X. Gu, H. Pan, Y. Liang, R. Castillo, D. Yang, D. Choi, E. Castillo, A. Majumdar, T. Guerrero, and S. B. Jiang, "Implementation and evaluation of various demons deformable image registration algorithms on a GPU," Physics in Medicine and Biology, vol. 55, no. 1, pp. 207–219, Jan. 2010.

[22] S. Klein, M. Staring, K. Murphy, M. A. Viergever, and J. P. W. Pluim, "elastix: A toolbox for intensity-based medical image registration," IEEE Transactions on Medical Imaging, vol. 29, no. 1, pp. 196–205, Jan. 2010.

[23] W. Huang, H. Yang, X. Liu, C. Li, I. Zhang, R. Wang, H. Zheng, and S. Wang, "A coarse-to-fine deformable transformation framework for unsupervised multi-contrast MR image registration with dual consistency constraint," IEEE Transactions on Medical Imaging, vol. 40, no. 10, pp. 2589–2599, Feb. 2021.

[24] G. Balakrishnan, A. Zhao, M. R. Sabuncu, J. Guttag, and A. V. Dalca, "VoxelMorph: A learning framework for deformable medical image registration," IEEE Transactions on Medical Imaging, vol. 38, no. 8, pp. 1788–1800, Aug. 2019.

[25] H. Xiao, R. Ni, S. Zhi, W. Li, C. Liu, G. Ren, X. Teng, W. Liu, W. Wang, Y. Zhang, H. Wu, H.-F. V. Lee, L.-Y. A. Cheung, H.-C. C. Chang, T. Li, and J. Cai, "A dual-supervised deformation estimation model (DDEM) for constructing ultra-quality 4D-MRI based on a commercial low-quality 4D-MRI for liver cancer radiation therapy," Medical Physics, vol. 49, no. 5, pp. 3159–3170, Feb. 2022.

[26] A. Ranjan and M. J. Black, "Optical flow estimation using a spatial pyramid network," in 2017 IEEE Conference on Computer Vision and Pattern Recognition (CVPR), 2017, pp. 2720–2729.

[27] N. Gunnarsson, J. Sjölund, and T. B. Schön, "Learning a deformable registration pyramid," in Segmentation, Classification, and Registration of Multi-modality Medical Imaging Data, Cham, 2021, pp. 80–86.

[28] T. N. van de Lindt, M. F. Fast, W. van den Wollenberg, J. Kaas, A. Betgen, M. E. Nowee, E. P. Jansen, C. Schneider, U. A. van der Heide, and J.-J. Sonke, "Validation of a 4D-MRI guided liver stereotactic body radiation therapy strategy for implementation on the MR-linac," Phys Med Biol, vol. 66, no. 10, May 2021.

[29] G. Li, J. Wei, M. Kadbi, J. Moody, A. Sun, S. Zhang, S. Markova, K. Zakian, M. Hunt, and J. O. Deasy, "Novel super-resolution approach to time-resolved volumetric 4-Dimensional magnetic resonance imaging with high spatiotemporal resolution for multi-breathing cycle motion assessment," International Journal of Radiation Oncology*Biology*Physics, vol. 98, no. 2, pp. 454–462, Jun. 2017.

[30] K. Sun and S. Simon, "A Resolution Enhancement Plug-in for Deformable Registration of Medical Images," p. 10.

[31] I. Goodfellow, J. Pouget-Abadie, M. Mirza, B. Xu, D. Warde-Farley, S. Ozair, A. Courville, and Y. Bengio, "Generative adversarial nets," in Advances in Neural Information Processing Systems, 2014, vol. 27.

[32] O. Ronneberger, P. Fischer, and T. Brox, "U-net: Convolutional networks for biomedical image segmentation," in Medical Image Computing and Computer-Assisted Intervention - MICCAI 2015, Cham, 2015, pp. 234–241.

[33] M. Jaderberg, K. Simonyan, A. Zisserman, and K. Kavukcuoglu, "Spatial transformer networks," in Advances in Neural Information Processing Systems, 2015, vol. 28.

[34] P. Kancharla and S. S. Channappayya, "Improving the visual quality of generative adversarial network (GAN)-generated images using the multi-scale structural similarity index," in 25th IEEE International Conference on Image Processing (ICIP), 2018, pp. 3908–3912.

[35] Z. Wang, E. P. Simoncelli, and A. C. Bovik, "Multiscale structural similarity for image quality assessment," in The Thrity-Seventh Asilomar Conference on Signals, Systems & Computers, 2003, vol. 2, pp. 1398-1402 Vol.2.

[36] Y. Liu, F.-F. Yin, Z. Chang, B. G. Czito, M. Palta, M. R. Bashir, Y. Qin, and J. Cai, "Investigation of sagittal image acquisition for 4D-MRI with body area as respiratory surrogate," Medical Physics, vol. 41, no. 10, p. 101902, 2014.

[37] J.-P. Thirion, "Image matching as a diffusion process: An analogy with Maxwell's demons," Medical Image Analysis, vol. 2, no. 3, pp. 243–260, Sep. 1998.

[38] A. Hore and D. Ziou, "Image quality metrics: PSNR vs. SSIM," in 20th international conference on Pattern Recognition, 2010, pp. 2366–2369.

[39] S. Zhi, M. Kachelrieß, and X. Mou, "High-quality initial image-guided 4D CBCT reconstruction," Medical Physics, vol. 47, no. 5, pp. 2099–2115, 2020.

[40] B. Huang, H. Xiao, W. Liu, Y. Zhang, H. Wu, W. Wang, Y. Yang, Y. Yang, G. W. Miller, T. Li, and J. Cai, "MRI super-resolution via realistic downsampling with adversarial learning," Phys. Med. Biol., vol. 66, no. 20, p. 205004, 2021.

[41] S. Zhi, M. Kachelrieß, F. Pan, and X. Mou, "CycN-Net: A convolutional neural network specialized for 4D CBCT images refinement," IEEE Transactions on Medical Imaging, vol. 40, no. 11, pp. 3054–3064, Nov. 2021.

[42] S. Zhi, M. Kachelrieß, and X. Mou, "Spatiotemporal structure-aware dictionary learning-based 4D CBCT reconstruction," Medical Physics, vol. 48, no. 10, pp. 6421–6436, 2021.

[43] A. Ammar, O. Bouattane, and M. Youssfi, "Automatic spatio-temporal deep learning-based approach for cardiac cine MRI segmentation," in Networking, Intelligent Systems and Security, Singapore, 2022, vol. 237, pp. 59–73.